\documentclass[a4paper,fleqn,usenatbib]{mnras}

\usepackage{graphicx}	
\usepackage{multirow}
\usepackage{hyperref}
\usepackage{xcolor}

\usepackage{times}
\usepackage{amssymb,amsmath}

\def\lsim{ \lower .75ex\hbox{$\sim$} \llap{\raise .27ex \hbox{$<$}} }
\def\gsim{ \lower .75ex \hbox{$\sim$} \llap{\raise .27ex \hbox{$>$}} }

\title[BL Lac neutrino candidates] 
{A multiwavelength view of BL Lacs neutrino candidates
}

\author[Righi et al.]
{C. Righi$^{1,2,3}$\thanks{E--mail: chiara.righi@brera.inaf.it}, F. Tavecchio$^2$, L. Pacciani$^4$ \\
$^1$ Universit{\`a} degli Studi dell'Insubria, Via Valleggio 11, I-22100 Como, Italy\\
$^2$ INAF -- Osservatorio Astronomico di Brera, via E. Bianchi 46, I--23807 Merate, Italy\\
$^3$ INFN -- Sezione di Genova, Via Dodecaneso 33, I-16146 Genova, Italy\\
$^4$ Istituto di Astrofisica e Planetologia Spaziali - Instituto Nazionale di Astrofisica (IAPS-INAF), Via Fosso del Cavaliere, 100 - I-00133
Rome (Italy)
}
\begin{document}



\maketitle

\begin{abstract} 
The discovery of high-energy astrophysical neutrinos by IceCube kicked off a new line of research to identify the electromagnetic counterparts producing these neutrinos. Among the extragalactic sources, Blazars are promising candidate neutrino emitters. Their structure, with a relativistic jet pointing to the Earth, offers a natural accelerator of particles and for this reason a perfect birthplace of high energy neutrinos. A good characterisation of the spectral energy distribution of Blazars can improve the understanding of their physical composition and the emission processes involved. Starting from our previous works, we select those BL Lac in spatial correlation with IceCube events. We obtain a sample of 7 sources and we start an observational campaign to have a better characterisation of the synchrotron peak. During our analysis a new source, namely TXS0506+056, has been added because of its position inside the angular uncertainty of a muon track event detected by IceCube. TXS0506+056, was in a high-state during the neutrino event and we will consider it as benchmark to check the properties of the other sources of the sample during the related neutrino detection. 
We obtain a better characterisation of the SED for the sources of our sample. A prospective extreme Blazar, a very peculiar low synchrotron peak (LSP) source with a large separation of the two peaks and a \textit{twin} of TXS0506+056 come up. We also provide the $\gamma$-ray light curve to check the trend of the sources around the neutrino detection but no clear patterns are in common among the sources. 
   \end{abstract}


\section{Introduction}

The recent detection of gravitational waves together with the discovery, few years ago, of an extraterrestrial component of high-energy neutrinos, inaugurated the era of multimessenger astrophysics. In particular, the IceCube detection of a still unresolved high-energy (above $\sim 60$ TeV to 2.8 PeV) neutrino diffuse emission (Aartsen et al. 2013, Aartsen et al. 2016) reveals the presence of astrophysical sources hosting hadrons at energies up to \textbf{$\sim 60$} PeV, possibly connected with the still mysterious sources of ultra-high energy cosmic rays.

Despite the low number of detected events ($\sim$ 85 since 2010), their distribution on the sky clearly excludes a purely galactic population of sources, leaving the possibility of a combination of galactic and extragalactic sources, as advocated by Palladino et al. (2016) and Palladino \& Winter (2018).

The basic ingredient required to produce high-energy neutrinos is a population of high-energy protons colliding  against matter ($pp$) or radiation ($p\gamma$). Both reactions produce charged pions, which quickly decay in electrons and neutrinos through the chain $\pi^{\pm} \rightarrow \mu^{\pm} + \nu_\mu \rightarrow e^\pm + \nu_e +2\nu_\mu$\footnote{We will not distinguish between $\nu$ and $\bar{\nu}$.}. The  neutrinos resulting from parent protons of energy $E_{\rm p}$ are characterized by an energy $E_{\nu }\approx E_{\rm p}/20$. The detection of $\sim 0.5-5$ PeV neutrinos therefore implies the presence of cosmic rays with energy in the range $10^{16}-10^{17}$ eV. The $pp$ channel is expected in region full of gas such as: galactic regions (Ahlers \& Halzen 2015), star forming galaxies (Tamborra et al. 2014, Loeb and Waxman 2006), low-power radio galaxies (Tavecchio et al. 2018), or galaxy clusters (Murase \& Beacom 2013; Zandanel et al. 2014). On the other hand, for other extragalactic sources, such as relativistic jets (in which we considered both active galactic nuclei, AGNs, and gamma ray burst, GRB) (see for examples Mannheim et al. 1993, Waxman \& Bahcall 1997; Tamborra et al. 2015), the jet density is expected to be too low and the most efficient interaction is expected to be the $p\gamma$.

Among active galactic nuclei (AGN), blazars (Romero et al. 2017) are often considered potential neutrino emitters  (Murase et al. 2014; Dermer et al. 2014) because their jets are thought to offer suitable conditions to accelerate the required high-energy protons (Kazanas \& Ellison 1986, Biermann \& Strittmatter 1987, Mannheim et al. 1991). Blazars belong to a subclass of AGN hosting a relativistic jet pointing at the Earth. The spectral energy distribution (SED) of this class is dominated by the relativistically beamed non-thermal emission of the jet with the characteristic "double hump" shape. The observed emission, predominantly originating in the jet, displays strong variations at all wavelengths and, due to the intense emission in the $\gamma$-ray band, they are the most numerous extragalactic $\gamma$-ray sources (Ajello et al. 2015). The first bump of the SED, peaking between the IR and the soft X-ray band, is due to synchrotron emission of relativistic electrons inside the jet. For the second bump, peaking in the $\gamma$-ray band, there are two main scenarios: the leptonic one ascribes the origin of this bump to the inverse Compton emission of the same electrons that generate the first bump (e.g. Ghisellini et al. 1998), while the synchrotron emission of protons or photo-meson reactions are the main mechanisms in the hadronic scenario (Aharonian 2000, Mucke et al. 2004). 

Depending on the luminosity and the presence of broad emission lines in the optical spectrum, blazars can be divided in two subclasses: Flat Spectrum Radio Quasars (FSRQs) and BL Lac objects (Urry \& Padovani 1995). The main differences between the two groups likely stem from the different regime of the accretion flow in these sources, which regulates the presence of the broad line region (BLR) (see e.g. Ghisellini et al. 2010, Sbarrato et al. 2012, 2014). High-accretion rates and the concomitant presence of a BLR are the main features determining the strong luminosity in the $\gamma$-ray band. From this perspective FSRQ could be a good neutrinos emitter candidates (Kadler et al. 2016). However there are a couple of difficulties against the idea that FSRQ are the main contributors to the neutrino flux observed by IceCube. One of these is given by Murase \& Waxmann (2016). Their point is based on the fact that the non-detection of neutrino multiplets by IceCube can be used to constrain the power and the cosmic density of potential sources. A population of powerful and  rare sources as FSRQ can be already excluded by current data if they have to account for the entire neutrino diffuse emission.  
 
Focusing on BL Lac objects, Padovani et al. (2016) (hereafter P16), showed a hint for a spatial correlation between the {\it highly peaked} BL Lac of the Second Catalog of Hard Fermi-LAT Sources (2FHL; Ajello, 2016) and a sample of IceCube events (both high-energy starting events, HESE, and tracks). Highly peaked BL Lac objects, or so called high synchrotron peak (HSP), are the subset of BL Lacs for which the maximum of the synchrotron peak occurs above a frequency of $10^{15}$ Hz. These sources are also the most abundant blazars detected in the TeV band. For these sources, Tavecchio et al. (2014) suggested that efficient neutrino emission can occur due to the possible structure of the relativistic jet, previously suggested in Ghisellini et al. (2005). In this scenario protons inside the fast jet core can interact with photons produced in a slower external layer triggering the photo-meson reaction. A possible association of an HESE IceCube event with and HSP was suggested by Padovani \& Resconi (2014) and Lucarelli et al. (2017). Inspired by these results, Righi et al. (2017) assumed a linear correlation between the neutrino flux $F_\nu$ and the $\gamma$-ray flux $F_\gamma$ of the BL Lac objects of the 2FHL and inferred the expected neutrino rate for each source.
In this framework, the $\gamma$-ray emission is thought to be dominated by the inverse Compton emission from the relativistic electrons. The $\gamma$-ray emission associated to the neutrino production (through the $\pi^0\to \gamma \gamma$ decay) is assumed to be subdominant and, after internal reprocessing, is expected to leave the source as a low-level flat component (e.g. Zech et al. 2017). This relation ($F_\nu \propto F_\gamma$, see equation 7 in Righi et al. 2017) is based on the assumption that both relativistic protons (producing neutrinos by $p\gamma$ reaction) and electrons (producing $\gamma$-rays) interact with the same photons of the layer.

A quite strong support to the idea that a a fraction of the neutrino flux is associated to BL Lacs comes from the 
recent possible association between a muon track event with an exceptionally good reconstructed direction and the active BL Lac TXS 0506+056 (Kopper \& Blaufuss 2017, Tanaka et al. 2017, Mirzoyan for the MAGIC Collaboration 2017). 
Many lepto-hadronic models are proposed to explain the neutrino emission from this source (Gao et al. 2018, Keivani et al. 2018, Murase et al. 2018, Cerruti et al. 2018, Ansoldi et al. 2018). In particular in Ansoldi et al. (2018) the detection is explained by the spine-layer model already proposed in Tavecchio et al. (2014).

To further investigate the hypothesis of BL Lacs as sources of neutrino events, we started a program aimed at obtaining a better multiwavelength characterization of the emission properties of these sources, and their modelling. First of all, we define a sample of 2FHL BL Lacs potentially associated to IceCube events. Then, we complemented very sparse existing MW data, with observations with the Neil Gehrels \textit{Swift} Observatory (hereafter \textit{Swift}) for three candidates of our sample, and with REM campaigns for two others sources. The final datasets allowed us to assemble (non-simultaneous) SED for the sources.
We describe our sample in Sect. 2. Data reduction and analysis is reported in Sect. 3 and in Sect 4 we describe and discuss the SED. Finally we discuss our results in Sect. 5. 

Throughout the paper, the following cosmological  parameters are assumed: $H_0=70$ km s$^{-1}$ Mpc$^{-1}$, $\Omega_{\rm M}=0.3$, $\Omega_{\Lambda}=0.7$.  We  use the notation $Q=Q_X \, 10^X $ in cgs units.

\section{Selection of the BL Lac neutrino candidates}

Following the results of P16 and Righi et al. (2017) mentioned in the previous section, we would like to assemble a sample of high-energy emitting BL Lacs (i.e. BL Lac detected above 50 GeV) to be correlated with the neutrino events. The best catalogue including this type of objects is the 2FHL catalogue. In fact, this catalogue consists of all sources detected above 50 GeV from the \textit{Fermi} satellite. Even if the most recent Fermi catalogue, the 3FHL, comprises more sources (711 instead of the 193 BL Lacs of the previous one), it includes all the sources detected at lower energies (above 10 GeV). For this reason we consider the 2FHL catalogue more suitable to select high-energy emitting BL Lacs.

\begin{figure*}
\hspace{-0.5 cm}
  \includegraphics{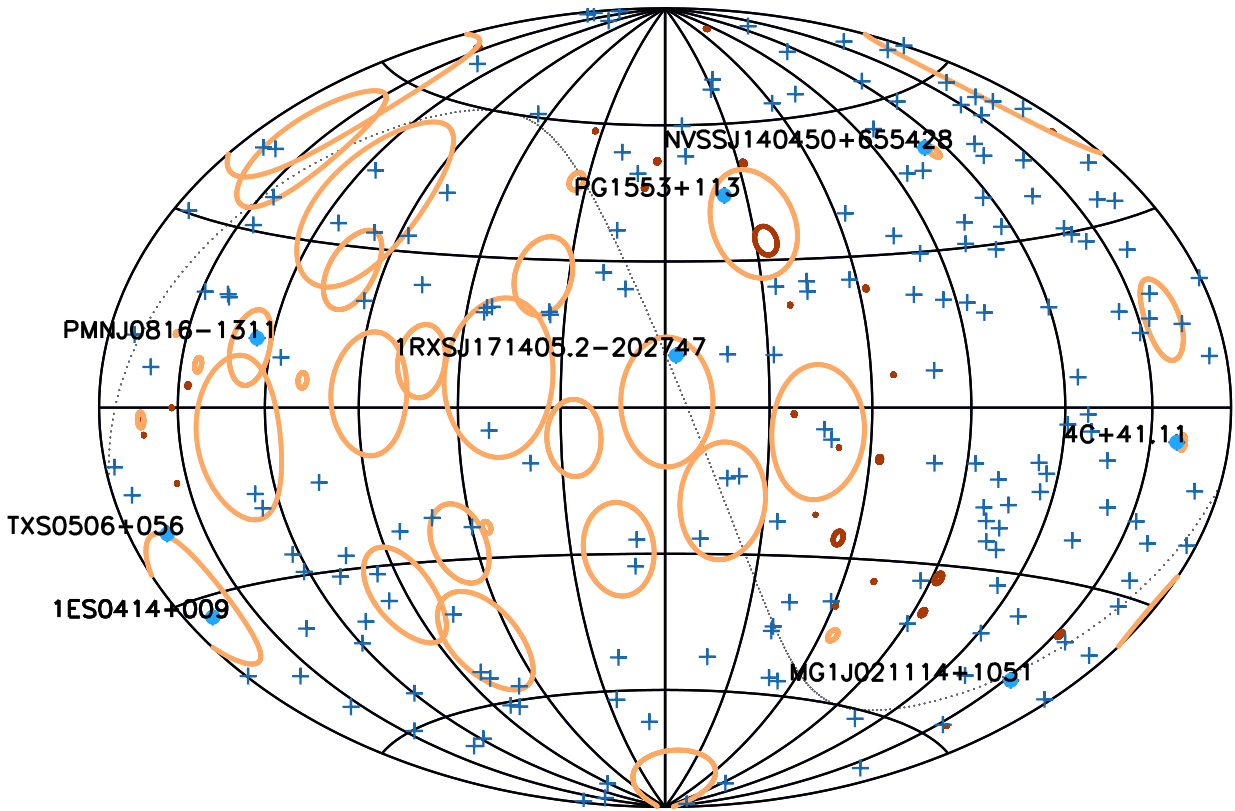}
  \caption{Sky map in galactic coordinates reporting the reconstructed direction of the neutrinos detected by IceCube.The dotted black line is the equatorial line, the orange circles corresponds to the angular uncertainty associated to 30 HESE events from P16 and the red dots indicates the direction of 29 muon tracks taken from Aartsen et al. 2016. The light blue crosses show the position of the 2FHL BL Lacs objects. We also indicate the sources in our sample (Table 2) with blue dots.}
  \label{fig:map}
\end{figure*}

To create a sample of BL Lacs belonging to the 2FHL catalogue and investigate a spatial correlation with a neutrino event, we use the list of neutrino events reported in P16. For the HESE events, P16 used the list provided by the IceCube Collaboration in Aartsen et al. (2014), including the events  recorded during the period 2010-2012. To reduce the background by atmospheric neutrino events they selected only the events with a reconstructed energy, $E_\nu \geq 60$ TeV. Moreover, to limit the number of counterparts, only the events with 
 angular uncertainty $\leq 20^\circ$ have been used. For the tracks, P16 considered the list given in Aartsen et al. (2015). For these tracks they assumed an average angular uncertainty of $0.4^\circ$, except for the $2.6$ PeV event, for which the median angular error is $0.27^\circ$ as reported in Schoenen \& Raedel (2015). Since a recent release of muon tracks events (from the northern hemisphere) is given in Aartsen et al. (2016), we combine the HESE list by P16 with this more recent list of tracks. 
The position and the corresponding uncertainty of the neutrino events included in our sample are reported in the sky map shown in Fig.\ref{fig:map} (HESE: orange circles; muon tracks: red circles), together with the 193 BL Lac of the 2FHL (blue crosses).

We selected all the BL Lacs whose positions lie within the (large) angular uncertainty of the HESE events. We list all the BL Lacs found to satisfy the above selection criteria in Table 1. 
For the  track events, instead, we choose to consider significant any case in which there is a BL Lac at a distance less than 2.5$^\circ$ from the reconstructed centroid of the neutrino direction. We chose this value, which for some events is larger then the 90\% C.L. angular uncertainty provided by IceCube Collaboration in Aartsen et al. (2016), to account for systematic differences between the reconstructed direction reported in the two different lists released by the IceCube Collaboration (respectively Aartsen et al 2015 and Aartsen et al. 2016). Note also than in both lists the angular errors are statistical errors only and do not include systematics. Then it is reasonable to consider a larger angular uncertainty.
Due to the large uncertainty associated to the reconstructed position of HESE, for 12 of these events we found more than one BL Lac inside the error circle of a single event. Given the ambiguity on the potential candidates, in this exploratory work we only consider the HESE events with only one association. In fact, bearing in mind that the aim of this work is to study a clean sample of candidate neutrino emitting BL Lacs, we considered only the events in one-to-one correspondence with a only one BL Lac.

\begin{table}
\centering
\begin{tabular}{clc}
\hline
ID $\nu$ & Source name & Class\\
\hline
\multirow{5}{*}{9}&RXJ0950.2+4553 & ISP\\
    &Ton1015 & HSP\\
    &Ton0396 & HSP\\
    &1H1013+498 & HSP\\
    &Mkn421 & HSP\\
 \hline
 \multirow{3}{*}{11}& RXJ1022.7-0112 & HSP\\
 &PMNJ0953-0840 & HSP\\
 &NVSSJ102658-174858 & HSP\\
 \hline
 \multirow{2}{*}{12}& PKS2005-489 & HSP\\
 &PMNJ1936-4719 & HSP\\
\hline
14 & 1RXSJ171405.2-202747 & HSP\\
\hline
17 & PG1553+113 & HSP\\
\hline
\multirow{2}{*}{19} & 1ES0505-546 & HSP\\
&1RXSJ054357.3-55320 & HSP\\
\hline
\multirow{3}{*}{20}&RBS0351 & HSP\\
      &PKS0229-581 & ISP\\
      &PKS0352-686 & HSP\\
 \hline
\multirow{3}{*}{22}&PMNJ1921-1607 & HSP\\
      &1H1914-194 & HSP \\
      &1RXSJ195815.6-30111 & HSP\\
 \hline
\multirow{2}{*}{26}&Ton0396 & HSP\\
      &MG1J090534+1358 & HSP\\
 \hline 
27 & PMNJ0816-1311& HSP\\
\hline
\multirow{2}{*}{30}&PMNJ0810-7530 & ISP\\
&PKS1029-85 & HSP\\
\hline
\multirow{2}{*}{33}&RXJ1931.1+0937 & HSP\\
   &1RXSJ194246.3+10333 & HSP\\
\hline
\multirow{4}{*}{35}&1RXSJ135341.1-66400 & HSP\\
&MS13121-4221 & HSP\\
&1RXSJ130737.8-42594 & HSP\\
&1RXSJ130421.2-43530 & HSP \\
\hline
\multirow{2}{*}{39}&TXS0628-240 & HSP\\
&PMNJ0622-2605 & HSP\\
\hline
41 & 1ES0414+009 & HSP\\
\hline
\multirow{2}{*}{51}&87GB061258.1+570222 & LSP\\
&GB6J0540+5823 & HSP\\
\hline
\end{tabular}
\label{Table:all}
\caption{List of all BL Lacs of the 2FHL catalogue in spatial correlation with a HESE neutrino event detected by IceCube and in the list of P16. As expected the majority are HSP, defined as BL Lacs with the synchrotron peak $\nu_S > 10^{15}$ Hz. }
\end{table}

\begin{table*}
\centering
\begin{tabular}{l|cc|cc|c}
\hline
Source name & $\alpha$ & $\delta$  & $z$ &$A_B$& $\nu$\\
& (J2000) & (J2000) & & & ID\\
\hline
\multicolumn{6}{c}{Single 2FHL BL Lac inside the angular uncertainty of the HESE events} \\
\hline
PMNJ0816-1311 & 124.113 & -13.197 & $>0.288^*$ & 0.296 &27$^a$\\
\textbf{1RXSJ171405.2-202747} & 258.521 & -20.463 & - &1.579& 14$^a$\\
\hline
\multicolumn{6}{c}{2FHL BL Lac with a distance max of 2.5$^\circ$ from a $\nu_\mu$} \\
\hline
\textbf{4C+41.11} & 65.983 & 41.834 & - & 2.665 & 13$^a$\\
\textbf{NVSSJ140450+655428} & 211.206 & 65.908 & 0.363 &0.049& 47$^a$\\
MG1J021114+1051 & 32.804 & 10.859 & 0.200 & 0.539 & 23$^b$\\
TXS 0506+056 & 77.358 &  5.693 & 0.336 & 0.392 & IC170922A$^c$\\
\hline
\end{tabular}
\label{Tableinfogen}
\caption{List of candidates neutrino sources studied in this work. For each source the equatorial (J2000) coordinates are reported (in degrees), the redshift, the $A_B$ extinction coefficient from  Schlafly \& Finkbeiner (2011) (recalibration 
of the Schlegel, Finkbeiner \& Davis (1998) infrared-based dust map) and the ID of neutrino detected by IceCube. The neutrino ID is taken from: $a$: Aartsen et al. 2014, $b$: Aartsen et al. 2015, $c$: IceCube Collaboration et al. 2018. Bold face characters identify the name of those sources for which we obtained dedicated \textit{Swift} pointings.
$*$: see Pita et al. (2014).}
\end{table*}

For the track events, we found one case in which the position of the associated BL Lac lies within the (small) angular uncertainty, namely MG1J021114+1051. As said above, to be conservative, we decided to include also two other sources, for which the distance from the corresponding neutrino position is less than 2.5$^\circ$. The total number of selected sources is therefore 7. Two of them, 1ES0414+009 and PG1553+113, are TeV sources already well studied in literature (Raiteri et al. 2017). Therefore in the following we only focus our attention on the remaining poorly known 5 sources, whose properties are listed in Table \ref{Tableinfogen}.  
During the preparation of this paper, the event IC170922A/TXS0506+056 occured. We obtained observation time at REM telescope (see next Section for details) and we added this BL Lac to our sample. In fact TXS0506+056 fully complies with our selection criteria (a source belonging to the 2FHL catalogue in the error circle of an IceCube event). Since this is the most plausible association observed so far, in our study we can use TXS 0506+056 as a benchmark case to discuss the other potential candidates (IceCube Collaboration at al. 2018). We note that the archival search of the past IceCube data revealed $13 \pm 5$  low energy muon neutrinos coming from the same region of TXS 0506+056 in the sky on a time scale of five months (IceCube-Collaboration 2018). These events were not present on the past lists released by the Collaboration.

An important point to note is that not all selected sources are HSP, the BL Lac subclass favored by the P16 analysis. In fact, even TXS 0506+056, whose synchrotron component peaks in the optical band, is classified as an intermediate synchrotron peak (ISP, $10^{14}$Hz$<\nu_S<10^{15}$Hz). Bold face is used for the sources for which we requested dedicated \textit{Swift} observations.
For MG1J021114+1051 and TXS 0506+056 we also obtained optical and IR observations with the Rapid Eye Mount (REM) telescope. For the other sources we only used archival data. The source TXS 0506+056 will be discussed in detail in the next section. 

 \section{Data analysis}
 
In the following, we describe the analysis performed on the \textit{Swift}/XRT,  \textit{Swift}/UVOT, REM and \textit{Fermi}/LAT data.

\subsection{REM data}
The Rapid Eye Mount telescope (REM) is a 60-cm robotic telescope located at the ESO La Silla Observatory. It includes an optical camera with the Sloan filters g, r, i, z and a near-infrared camera equipped with J-H-K filters. In these bands we observed MG1 J021114+1051 and TXS 0506+056. Data reduction was carried out following the standard procedures, with the subtraction of an averaged bias frame dividing by the normalised flat frame. The photometric calibration was achieved by using the 2MASS and APASS catalogues. In order to minimise any systematic effect, we performed differential photometry with respect to a selection of non-saturated reference stars. Table 3 shows the observation period of these sources and the magnitude obtained at different filters.

\begin{table*}

\centering
\begin{tabular}{c|cccccc}
\hline
 Period &   \multicolumn{6}{c}{Filters}  \\
& J & H & K & g & r & i \\
\hline
\multicolumn{7}{c}{MG1J021114+1051}\\
\hline
  01Oct2016/25Nov2016           &  - & - & $17.088\pm0.010$ & $ 15.111\pm0.025$ & $14.578\pm0.022$ & $14.163\pm  0.032$\\
  \hline
  \multicolumn{7}{c}{TXS 0506+056}\\
  \hline
                 30Sep2017      &  $ 12.781\pm0.056$ & $11.945\pm0.035$ & $11.205\pm0.100$ & $15.013\pm0.026$ & $14.547\pm0.020$& $14.174\pm0.032$\\
                 01Oct2017       &  $12.632\pm0.042$ & $ 11.930\pm0.051$ & $11.061\pm0.064$ & $14.867\pm0.022$ & $14.361\pm0.022$& $14.037\pm0.033$\\
                 \hline
\end{tabular}
\label{TableREM}
\caption{Observation period and filter used for the observation with REM telescope.}
\end{table*}

  \begin{table*}
\centering
\begin{tabular}{l|cccccc}
\hline
Source name & \textit{b} & \textit{m2}  & \textit{u} & \textit{v} & \textit{w1} & \textit{w2}\\
\hline
1RXSJ171405.2-202747 & $19.55 \pm 0.32$ & $>20.93$ & $>19.41$ & $18.45 \pm 0.28$ & $ > 20.27$ & $> 21.13$ \\
4C+41.11 & $21.32 \pm 0.42$ & $> 21.45$ &$21.22 \pm 0.53 $ & $20.07 \pm 0.32$ & $ > 21.08$ & $> 21.69$ \\
NVSSJ140450+655428 & $> 20.18$ & $> 20.53$ & $> 19.78$ & $> 19.38$ & $ > 20.00$ & $> 20.84$\\
\hline
PMN J0816-1311 & $17.19 \pm 0.03$ & $16.24 \pm 0.03$ & $16.25 \pm 0.03$ & $16.82 \pm 0.04$ & $16.18 \pm 0.03$ & $16.29 \pm 0.03$ \\
MG1 J021114+1051 & $14.55 \pm 0.01$ & $14.46 \pm 0.02$ & $14.38 \pm 0.02$ & $14.02 \pm 0.02$ & $14.13 \pm 0.02$ & $14.58 \pm 0.01$ \\
TXS 0506+056$^h$& $15.06\pm0.02$&$14.46\pm0.02$ &$14.27\pm0.02$ &$14.61\pm0.02$ &$14.35\pm0.03$ &$14.58\pm0.02$ \\
TXS 0506+056$^l$& $15.74\pm0.04$&$15.42\pm0.08$ &$15.04\pm0.03$ &$15.24\pm0.04$ & $15.27\pm0.03$&$15.60\pm0.03$\\ 
\hline

\end{tabular}
\label{TableUVOT}
\caption{\textit{Swift}/UVOT observed magnitudes. Statistical uncertainties only are reported: systematic error is always lower than 0.03 mag. For TXS 0506+056 there are two states: $h$: high state of the source on 27/09/2017 (MJD: 58023.752), $l$: low state of the source on 25/07/2009 (MJD: 55037.512).}
\end{table*}

\begin{table*}
\centering
\begin{tabular}{l|cccccc}
\hline
Source name                    & Exp. time & $\Gamma (\beta)$                  &          $N_H$        &$E_p$       & $\chi^2_{\textit{red}} ($d.o.f.)  & F$_{0.3-10 \text{keV}}$\\
                                          &   [ks]         &                                   & [$10^{21}$ cm$^{-2}$] &                         & [$10^{-12}$ erg cm$^{-2}$ s$^{-1}$]\\
\hline
1RXSJ171405.2-202747  &   10.14       & $1.88\pm 0.1$  & 1.56                   &     --       &  2.2 (13)       &  $6.1 \pm 0.4$\\  
1RXSJ171405.2-202747  &   10.14       & $0.86^{+0.36}_{-0.32}$  & 1.56            &   --  $1.83^{0.32}_{-0.27}$              &  0.49 (12)       &  $5.0^{+0.3}_{-0.4}$\\        
      
4C+41.11                         &   27.70      & $ 1.578 \pm 0.103$  & 3.38                   &      --      &   0.772 (12)     & $0.64\pm0.55$\\
NVSSJ140450+655428  &   10.78      & $2.349 \pm 0.089$    & 0.171                &      --       & 1.083 (12)         & $1.44\pm 0.05$\\
\hline
PMNJ0816-1311              & 6.87        & $2.296 \pm 0.026$     & 0.81                   &     --      &   1.394 (128) & $19.97\pm 0.05$\\
MG1J021114+1051         &  18.98     & $2.176 \pm  0.027$      & 0.616                &      --       &  1.201 (121)         & $5.79\pm 0.11$\\
TXS 0506+056$^h$ & 4.947 & $2.606\pm0.089$& 1.11 &-- & 1.016 (21) &$ 3.07\pm0.25$ \\
TXS 0506+056$^l$ & 4.491 & $2.139\pm0.288$& 1.11 &-- & 0.282 (2) &$0.86\pm0.15$ \\
\hline
\end{tabular}
\label{TableXRT}
\caption{Results of the \textit{Swift}/XRT data analysis. For TXS 0506+056 there are two states: $h$: high state of the source on 27/09/2017 (MJD: 58023.752), $l$: low state of the source on 25/07/2009 (MJD: 55037.512). For all sources an absorbed power-law model provides a good representation of the spectrum. 1RXSJ171405.2-202747 source was modelled with an absorbed power-law (upper row) and a log parabolic fit (lower row), parametrized by $\beta$ and $E_p$.}
\end{table*}

\subsection{\textit{Swift}}

\textit{Swift} is a satellite equipped with several instruments (Burrows et al. 2005). For all sources listed in Tab. 2, we had snapshot observations for both optical/UV and X-ray data. Comparing the different observation we noticed low variability and then we sum all the observations to increment the signal to noise ratio. In particular, we asked and obtained observation time for three sources of our sample (the bold face reported in Table 2). The observations were performed in the period October 2016-July 2017. For the other sources instead we re-analysed the archival data. In particular MG1J021114+1051 were observed in the period March 2010-November 2011 (data were already published, see Chandra et al. 2014 for details) while PMN J0816-1311 was observed by \textit{Swift} in 2009.
\subsubsection{\textit{Swift}/UVOT data}
\begin{figure}
\hspace{-0.5 cm}
  \includegraphics[width=0.5\textwidth]{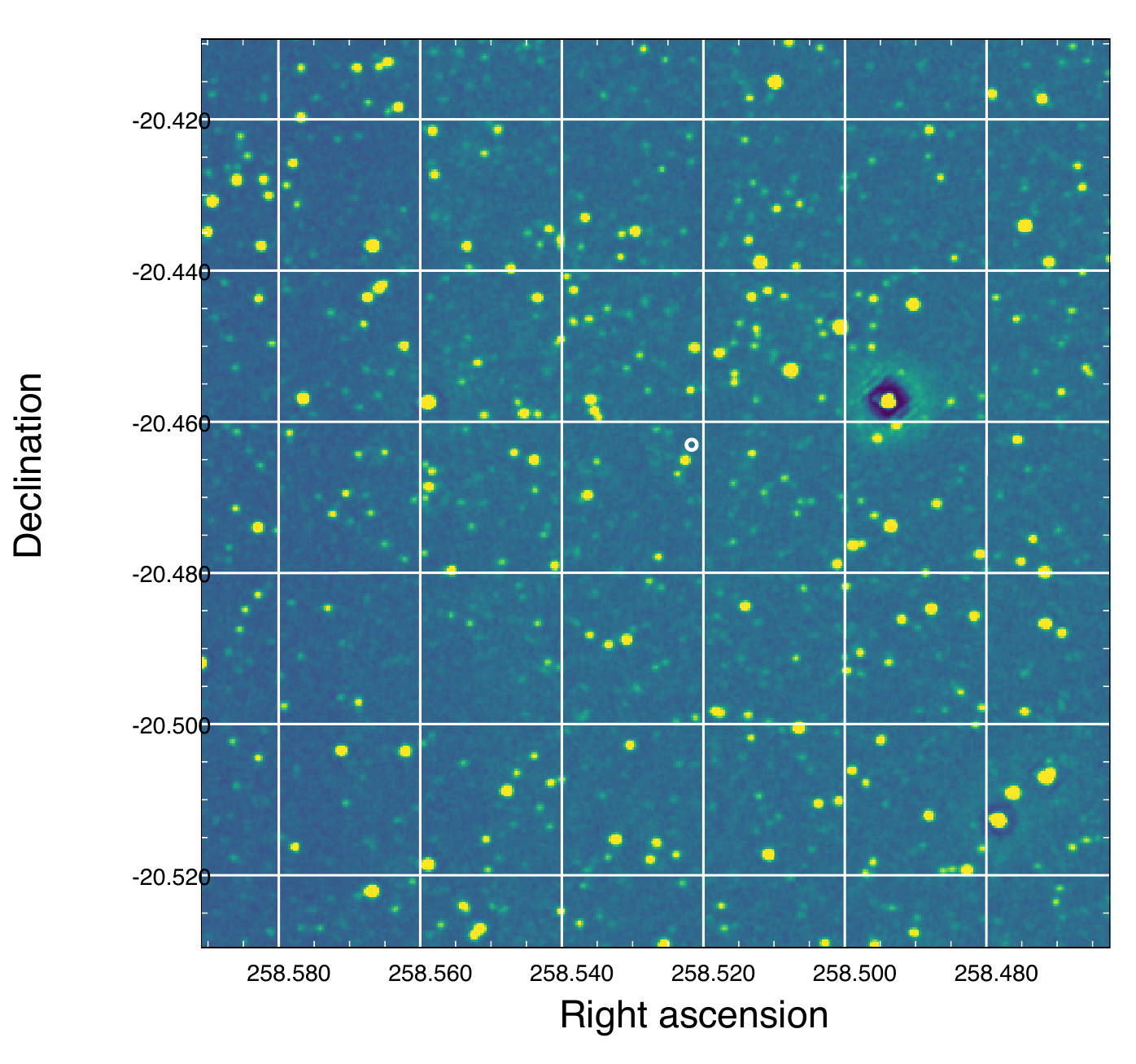}
  \caption{Position map of the source 1RXS J171405.2-202747 in the \textit{Swift}/UVOT B filter. The position of the source is highlighted by the white circle.}
  \label{fig:position1RXS}
\end{figure}

 The satellite \textit{Swift} includes a 30 cm diffraction-limited optical-UV telescope (UVOT) (Roming et al. 2005) equipped with six different filters that covered the $170-650$nm wavelength range, in a 17 arcmin $\times$ 17 arcmin FoV.
 From the High Energy Astrophysics Science Archive Research Center (HEASARC\footnote{\url{https://heasarc.gsfc.nasa.gov/docs/archive.html}}) data base we download the UVOT images in which our target sources were observed.  For all the sources the analysis was performed with the \texttt{fappend}, \texttt{uvotimsum} and \texttt{uvotsource} tasks\footnote{\url{https://heasarc.gsfc.nasa.gov/docs/software/lheasoft/}}. Due to the position of 1RXSJ171405.2-202747 full-stars field (see Fig.\ref{fig:position1RXS}) we perform a dedicated analysis. For the other sources we use a source region of 5 arcsec and the background was extracted from a source-free circular region with radius equal to 20 arcsec. The extracted magnitudes were corrected for Galactic extinction using the values of Schlegel et al. (1998), reported in the second to last column of Table \ref{Tableinfogen} and applying the formulae by Pei 1992 for the UV filters, and eventually were converted into fluxes following Poole et al. 2008.
 Table 4 reports the observed Vega magnitudes in the Swift/UVOT $v$, $b$, $u$, $m1$, $m2$, and $w2$ filters, together with statistical uncertainties. Systematic uncertainties are never greater than 0.03 mag and therefore dominated by statistical ones in the vast majority of cases.

\subsubsection{\textit{Swift}/XRT data}
\textit{Swift}/XRT (Burrows et al. 2005) data were analysed by using HEASOFT v6.20 software package. We analysed the spectra of the sources with XSPEC v.12.9.1 (Dorman \& Arnaud, 2001)  in order to extract the flux in the $0.3-10$ keV energy band and the photon index $\Gamma$, using the $\chi^2$ minimization.  For all sources an absorbed power-law model provides a good description of the spectrum. In all cases the fits are compatible with an absorption column, $N_\text{H}$, fixed to the Galactic value. Table 5 shows the best fit parameters.

\subsection{\textit{Fermi}/LAT data}
Fermi-LAT data analysis was performed using the Fermi Science Tools
(v10r0p5) and PASS8 response Functions (P8R2\_SOURCE\_V6).
Gamma-ray data  were selected running {\em gtselect} for SOURCE events class, collected within 20$^\circ$ from the source under investigation; the chosen zenith angle cut was 90$^\circ$. GTIs were prepared running {\em gtmktime} to select good quality data, collected during standard data taking mode. Livetime cubes were prepared taking into account the chosen zenith angle cut.\\ 

Gamma-ray light curves were produced in the energy range 0.3-100 GeV with a bin size of 4d and 16d for all sources. To cover possible active states lasting for several months, as observed in the case of TXS0506+056, we show the light curve in an interval of 300 days centred around the associated neutrino event. The flux reported for the chosen time-bins of the light curves is obtained with the standard unbinned likelihood analysis. The sources input files for the unbinned likelihood was prepared starting from the sources positions and spectral templates reported in the 3FGL catalog (Acero et al. 2015). For the investigated source, normalization and spectral parameters were allowed to vary. For sources within 10$^\circ$ from the investigated source, the normalization factor only  was allowed to vary, and all the spectral parameters were fixed to their catalog value. For sources outside 10$^\circ$ from the investigated source, the normalization and all the spectral parameters were fixed to their catalog value. 

\begin{figure*}
  \includegraphics[width=15cm, height=20.cm]{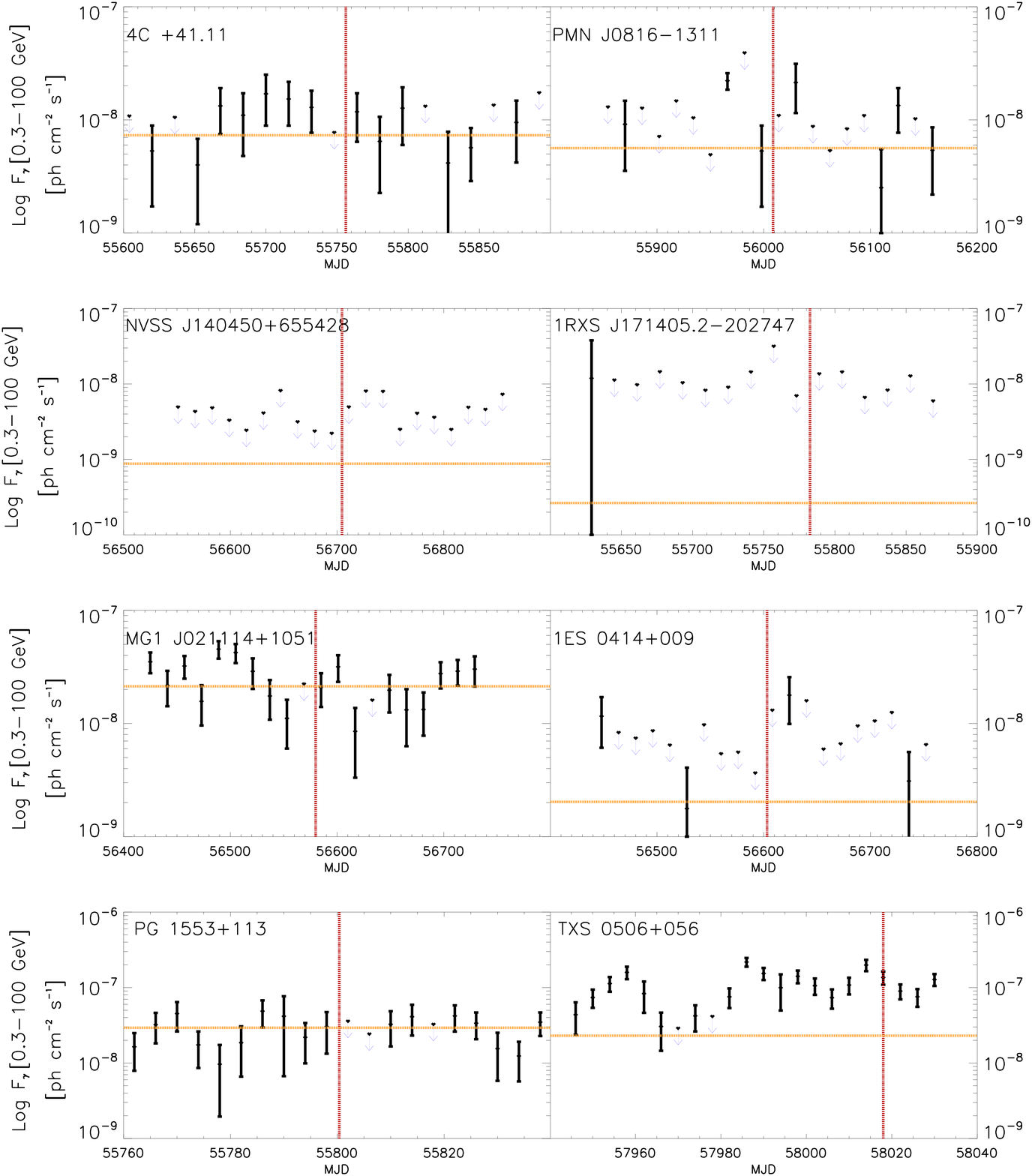} 
  \vspace*{-1.4cm}
  \caption{$\gamma$-ray light curve of all 8 candidate sources. The bin is 16days apart from PG1553+113 and TXS 0506+056 in which the bin is 4 days. The horizontal orange line represents the mean flux reported on the 3LAC catalogue. The data do not show flares in correspondence with the neutrino emission (red vertical line) however a discussion about the expected coincidence between a neutrino event and a $\gamma$-ray flare is in Section 5.}
  \label{fig:lc}
\end{figure*}

In Fig. \ref{fig:lc} we show the light curve for our sample of source (including PG1553+113 and 1ES0414+009). Due to the low flux, we show the 16 days bin light curve for all the source except for PG1553+113 and TXS0506+056 that are bright enough to have a good light curve with 4 days bin. The red vertical line shows the arrival time of the neutrino in spatial correlation with the source and the orange horizontal line is the mean flux of the source in the range 0.3-100 GeV reported in the 3FGL Fermi catalogue.

From Fig. \ref{fig:lc} it is clear that for the most of the sources there is no significant $\gamma$-ray activity at the time of the neutrino detection (except the case of TXS0506+056). The light-curve of TXS0506+056 is consisted with the one present in IceCube Collaboration et al. (2018), Padovani et al. (2018), Keivani et al. (2018) and Veritas Collaboration (2018).

\subsection{The case of 1RXSJ171405.2-202747}
The study of 1RXSJ171405.2-202747 needed of a careful analysis because of the position of the source. It is in fact very close to the galactic center and therefore in a region full of stars and other sources (see Fig.  \ref{fig:position1RXS}). For this reason we check carefully every data related to this source to be sure the effective association with our source.

Within the 3FGL gamma-ray catalog (Acero et al. 2015) the accuracy in the position of 3FGL J1714.1-2029 is 3.6 arcmin (95\% c.l.). 
1RXS J171405.2-202747 is identified as its X-ray counterpart.\\
At 2.0 arcmin from the $\gamma$-ray source there is a radio source (Condon et al. 1998):
NVSS J171405-202748 (with an accuracy on the position of radio source of 2.4 arcsec R.M.S.); while at 4.9 arcmin (just oustide the $\gamma$-ray error
circle) there is NVSS J171402-202525. NVSS J171357-203653 is at 7.5 arcmin,
NVSS J171442-202631 at 8.7 arcmin, all the other NVSS sources are more than 10
arcmin apart from the $\gamma$-ray source.\\
An X-ray source was observed and detected with Swift several
times at celestial coordinates: $\alpha$ = 17 14 05.4, $\delta$=-20$^\circ$ 27' 49'',
with an error of 3'', coincident with the position of
1RXS J171405.2-202747 and of NVSS J171405-202748. No X-ray counterpart is
found for NVSS J171402-202525.\\
In the following we will assume the detected \textit{Swift} source as the X-ray
counterpart of 3FGL J1714.1-2029; and NVSS J171405-202748 as the radio
counterpart of the $\gamma$-ray source.\\
There is a weak near IR counterpart for NVSS J171405-202748 found in the 2MASS
catalog (Skrutskie et al. 2006), with celestial coordinates $\alpha$=17 14 05.43, $\delta$=-20$^\circ$ 27' 49.09''
and positional error of 0.15 arcsec. A brighter NIR object  ($\alpha$ =17 14 05.44, $\delta$= -20$^\circ$ 27' 54.27''  ) is found at 6.1 arcsec
from  NVSS J171405-202748, just  outside the radio source error circle.
We will consider the first near IR source as the counterpart for NVSS J171405-202748.\\ 

Summing-up all \textit{Swift}/XRT observations, an absorbed power-law model does not fit
to the data (reduced $\chi^2$=2.2, see Table 5). A log-parabolic (Tramacere et al. 2007) model fit
to the data: Using the \emph{eplogpar} function ($F(E)=\frac{K}{E^2}10^{-\beta(log(\frac{E}{E_P})^2}$),
the estimated parameters (for a confidence level of 90\%) are: peak energy $E_P =  1.83^{+0.32}_{-0.27}$,
curvature term $\beta\ =\ 0.86 ^{+0.36}_{-0.32}$, unabsorbed flux (in the 0-3-10 keV energy range) $F\ =\ (5.0 ^{+0.3}_{-0.4})10^{-12}$ erg cm$^{-2}$ s$^{-1}$.
The $\chi^2$ is 5.9 for 12 degree of freedom, the null hypothesis probability is 0.92.\\

\section{Spectral Energy Distributions}

\begin{figure}
\hspace{-1.4cm}
  \includegraphics[width=12cm, height=13.cm]{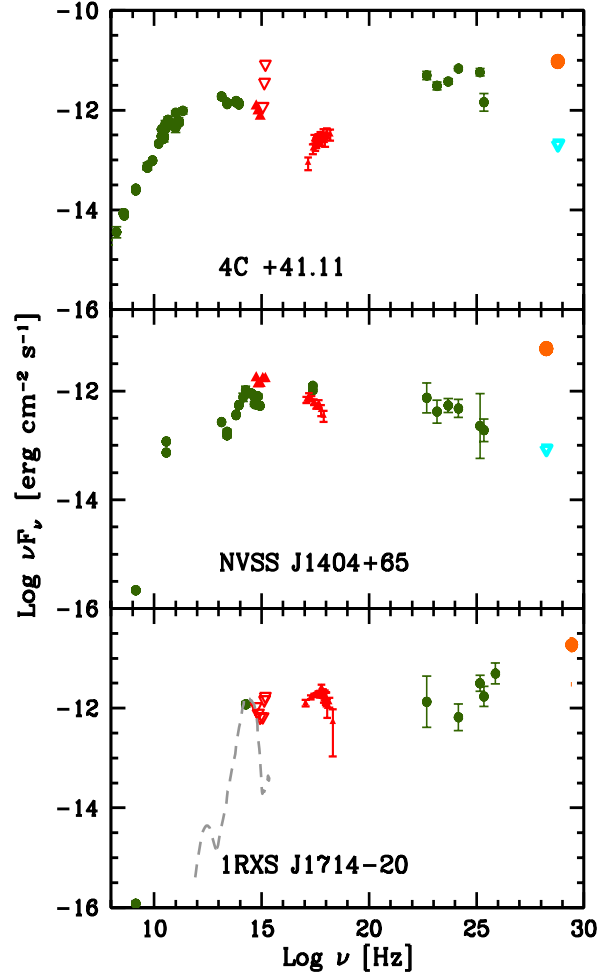} 
  \caption{Spectral energy distributions for three of BL Lac neutrino candidates. Green dots are archival data (by ASDC), red filled up-pointing triangle are \textit{Swift}/UVOT and \textit{Swift}/XRT data. \textit{Swift}/UVOT upper limits are indicated with red down-pointing triangle. Orange dots corresponds to the expected neutrino flux, assuming one neutrino in 7 years of observation by IceCube, and using the effective area at the energy of the neutrino associated with the BL Lac. Light-blue triangle is the neutrino flux calculated in Righi et al. (2017). Due to the declination of 1RXSJ1714-20 (below the equator, the neutrino flux calculated in Righi et al. (2017) is missing.}
  \label{fig:sed1}
\end{figure}

\begin{figure}
\hspace{-1.4cm}
    \includegraphics[width=12cm, height=13.cm]{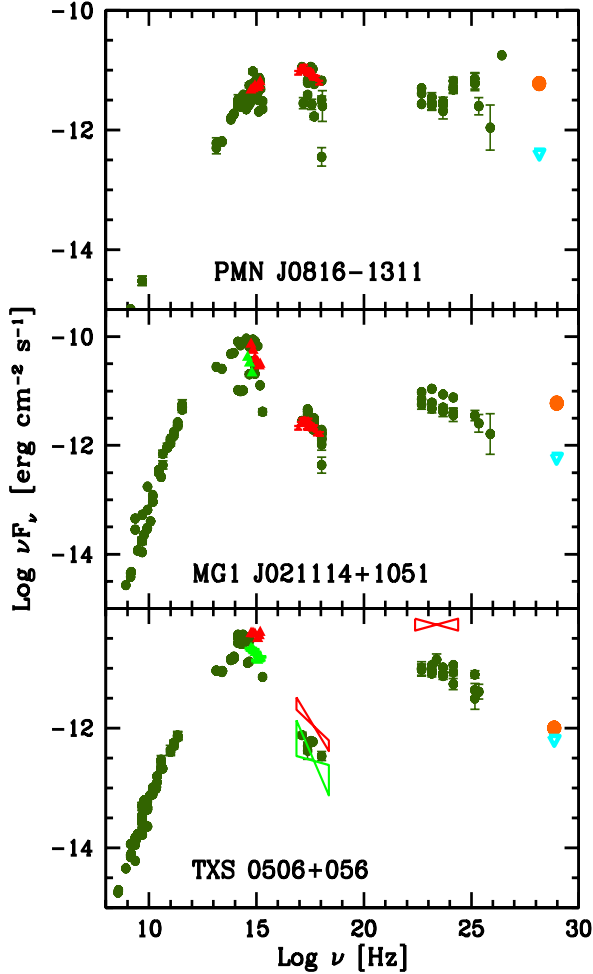} 
  \caption{Spectral energy distributions for three of BL Lac neutrino candidates. Green dots are archival data (by ASDC), red filled up-pointing triangle are \textit{Swift}/UVOT and \textit{Swift}/XRT data. \textit{Swift}/UVOT upper limits are indicated with red down-pointing triangle. Light green up-pointing triangle are REM data (for MG1 J021114+1051and TXS0506+056) and \textit{Swift}/UVOT estimation (only for TXS 0506+056). Orange dots corresponds to the expected neutrino flux, assuming one neutrino in 7 years of observation by IceCube, and using the effective area at the energy of the neutrino associated with the BL Lac. Light-blue triangle is the neutrino flux calculated in Righi et al. (2017).}
  \label{fig:sed2}
\end{figure}

The Spectral Energy Distributions (SED) of the 6 sources, built by using archival data (green) and the data described above, are shown in Figs. \ref{fig:sed1}-\ref{fig:sed2}.

We remark that the observational data from \textit{Swift} and REM are not simultaneous. Moreover \textit{Swift} spectra have been obtained from short snapshots performed over several months (see section 3.2). Furthermore, there are very few data during the neutrino detection. The SED can therefore only provide time average information and cannot be used for detailed modelling of the electromagnetic and neutrino output.


The SED display a large variety of shapes. In particular, two sources (PMN J0816-1311 and NVSS J1404+65) clearly belong to the HSP population, with a peak frequency of the synchrotron component above 10$^{15}$ Hz. MG1 J021114+1051 and TXS 0506+056 display a quite notable similarity and fulfil the criteria to be defined ISP. The SED of the remaining two sources have a less clear nature. 

As discussed above, the analysis of the data of 1RXS J1714-20 is complicated by its position on the sky, close to the galactic plane. In particular, the confusion introduced by the complexity of the field makes difficult to understand the correct association of some of the data found in literature.
For this reason we made a careful selection of the archival data. 
The concave X-ray spectrum from XRT, modelled with a log-parabolic fit (see section 3.5 for detail), suggests a peak around 1 keV. Such a large synchrotron peak frequency resemble a characteristic feature of the so-called extreme BL Lacs (e.g. Costamante et al. 2001, Bonnoli et al. 2015, Costamante et al. 2018). Besides a peak in the X-ray band, these peculiar sources display a quite hard gamma-ray continuum, often peaking in the TeV band. The optical band, instead, is dominated by the emission from the host galaxy. The data for 1RXS J1714-20 are consistent with both characteristics. The LAT data track a hard spectrum peaking above 100 GeV. The exceptional hardness of the spectrum is confirmed by the fact that this source belongs to the 2FHL (selection above 50 GeV) but it is absent in the 3FHL (selection above 10 GeV). Unfortunately, the description of the optical emission is poor. However, the UVOT upper limits together with the 2MASS datapoint are consistent with the emission from a typical elliptical host galaxy of BL Lac objects (for comparison, the dashed line reports the template for a giant elliptical by Silva et al. 2004).

The SED associated to 4C+41.11 is  puzzling. The archival and the UVOT data locate the maximum of the synchrotron peak in the IR band. The hard XRT spectrum suggests that the X-ray continuum is associated to the second bump, likely peaking in the LAT energy band.
The position of the synchrotron peak define 4C+41.11 as a LSP. However, the flat LAT spectrum (photon index $\approx 2$) is quite atypical for this class (Ackermann et al. 2015). The shape of this SED is quite similar to the case of AP Lib, another LSP with an unusually hard LAT spectrum. This particular SED is quite difficult to be reproduced with standard one-zone emission models (e.g. Tavecchio et al. 2010) and possibilities to overcome this problem include the addiction of other components, possibly from the large-scale jet (Hervet et al. 2015, Sanchez et al. 2015, Zacharias \& Wagner 2016), or the contribution of hadronic processes (Petropoulou et al. 2017).

The case of TXS0506+056 has raised the attention of the whole high-energy astrophysics community (Kopper \& Blaufuss 2017, Tanaka et al. 2017, Mirzoyan for the MAGIC Collaboration 2017). The facts that the source was in an high-state in the $\gamma$-ray band during the neutrino detection, that the event was a muon track event with a very good reconstructed direction (less than $1^\circ$) and the detection for the first time in the TeV band, make this event unique and particularly relevant. 
A full description of the neutrino event in spatial and temporal correlation with TXS0506+056 is given in IceCube Collaboration et al. (2018) in which a multiwavelength analysis is also reported, and a theoretical analysis of the low and the high state is given in Ansoldi et al. (2018), Cerruti et al. (2018), Gao et al. (2018), Keivani et al. (2018), Murase et al. (2018).
Paiano et al. (2018) showed the optical spectrum of the sources taken with the Gran Telescopio CANARIAS (GTC) with which, thanks to the emission lines of [OII],[OIII] and [NII], they attested a redshift of $z=0.3365 \pm 0.0010$.  Here we report both the high state, with data taken in the period 27/09/2017-01/10/2017, and the low state, data of 25/07/2009.

Together with the electromagnetic output, in Figs. \ref{fig:sed1}-\ref{fig:sed2} we also report the inferred level of the neutrino emission. In particular, the orange circles have been derived calculating the expected neutrino flux, $F_{\nu_c}$ required to have one neutrino detected during the seven year of operation of IceCube and assuming the energy estimated for that event. To this aim we use the declination-dependent effective area provided by Yacobi et al. (2014) for track events and the one performed in Niederhausen et al. (2015) for the HESE. The light blue triangle instead show the flux, $F_{\nu_{R17}}$ derived by using the model of Righi et al. (2017), which assumed that BL Lacs belonging to the 2FHL account for the entire observed neutrino diffuse emission and that for each source the neutrino flux is correlated to its $\gamma$-ray flux. For this reason the light-blue triangle are to consider upper-limits, because there are arguments suggesting that BL Lacs contribute only a fraction to the entire neutrino diffuse emission observed bu IceCube (e.g. Palladino \& Winter 2018, Aartsen et al. 2016, Aartsen et al. 2017) The fact that the brightest BL Lac sources of 2FHL catalogue are absent from our sample (such as Mkn421 or Mkn 501), suggests an overestimation of the flux $F_{\nu_{R17}}$. This raises a question about the neutrino emission from Mkn-like sources (see Righi et al. in prep.). Note that in Righi et al. (2017) we considered only the northern hemisphere, for this reason, for the 1RXS J 1714-20, we present only $F_{\nu_c}$.

The requirement to produce a sizeable neutrino emission, implies that a fraction of the electromagnetic output derives, at least, from the $\gamma$-rays and the pairs injected in the source after the decay of neutral and charged pions. To properly model these processes (in particular the associated electromagnetic cascades) one needs to fully implement all the processes as in e.g., Mannheim (1995) and Boettcher et al. (2013). However, the paucity of soft target photons provided by the synchrotron component alone, requires the existence of external sources, such as the photons from the accretion flow (Righi et al., in prep) or those envisioned in the spine-layer scenario (e.g. Tavecchio et al. 2014). 

\section{Discussion}
Following the idea that BL Lacs can be the emitters of high-energy neutrinos detected by IceCube, we started a observational campaign of a sample of candidates. From a list of 30 HESE +  29 muon tracks events respectively from P16 and Aartsen et al. (2016), and the BL Lac of the 2FHL catalogue of Fermi, we obtain a sample of 8 candidate neutrino BL Lacs spatially correlating with IceCube events. 
Two of the sources are very well-known high-energy emitting BL Lacs  detected also in the TeV band (PG1553+113 and 1ES0414+009). For the other six sources we obtained observations with REM and \textit{Swift} (optical, UV and X-ray band), to have a more accurate description of the synchrotron peak. Adding also archival data we derive the spectral energy distribution, that show a variety of shapes. As expected (since we started from 2FHL objects), the majority of sources are HSP, i.e. display a synchrotron peak at frequencies $\nu_S>10^{15}$Hz, but, over a total of 8 sources, 3 appear to belong to the LSP or ISP subclasses. Assuming the detection of only one neutrino in 7 years with IceCube, we calculate the expected muon neutrino energy flux ($F_E=N\cdot E/A_{\rm eff}t$, with $N=1$, $t=7$y and $A_{\rm eff}$ the muonic effective area at the specific declination and energy and $E$ the reconstructed neutrino energy), obtaining values in the range $10^{-12}<F_{\nu}< 10^{-11}$ erg cm$^{-2}$ s$^{-1}$. We also compare this values with the expected muon neutrino flux obtained in a previous work (Righi et al. 2017). The latter are systematically lower than those derived above assuming the detection of one neutrino in seven years. However,  in considering this result, it is important to keep in mind that these fluxes -- whose derivation assume, for instance, a constant flux of the sources (even if the large scale variability is one of the main characteristic of this class) and that this class is the unique emitter of the IceCube events -- are affected by large  uncertainties. 

To investigate the possibility that the neutrino emission is associated to a particularly active state of the sources we have derived the light curves in the LAT band. While in the case of TXS 0506+056 the neutrino detection (Sep. 2017) coincides with a long lasting active state starting in April 2017 (see IceCube Collaboration et al. 2018), none of the other sources show such a significant increase of activity close to or in correspondence of the epoch of the neutrino detection. Small amplitude variability possibly correlated with the neutrino detection occurred in MG1 J021114+1051, PMN J0816-1311 and in 1ES 0414+009. However the quality of the data prevent any conclusion. A dedicated analysis of the correlation between the LAT light curves and possible excesses recorded by IceCube around the position of these sources could be interesting. However, we note that a strict correlation between neutrino emission and $\gamma$-ray activity is  questionable also for TXS 0506+056, as proved by the potential neutrino emission found in 2014/2015 by Aartsen et al. (2018b) in coincidence with a rather quite gamma-ray state (Padovani et al. 2018). The modelling of the multimessanger SED of TXS 0506+056 shows that the $\gamma$-ray peak cannot be dominated by the radiation product of the photo-hadronic scenario. 
The low sensitivity of present neutrino detectors with the pronounced variability of the sources make difficult the assessment of correlations between $\gamma$-ray flares and the neutrino emission. 

A potential problem of the framework linking BL Lacs and the neutrino diffuse emission is represented by the absence of any clear association of neutrinos with the two brightest representative of the class, Mkn 421 and Mkn 501 (see also discussions in Aartsen et al. 2018b). In fact, there are no events  associated with Mkn 501, while Mkn 421 is only potentially associated to a cascade events whose reconstructed direction is characterized by a very large angular uncertainty (Padovani \& Resconi 2014, Petropoulou et al. 2015). The lack of events clearly correlated with these sources, after 7 years of activity by IceCube, raises doubts about the role of HSP as important neutrino emitters. Indeed, estimates based on the high-energy $\gamma$-ray flux as proxy (e.g., Righi et al. 2017) suggest that these two sources alone should provide $\sim 50\%$ of the entire muon neutrino emission attributable to BL Lacs. In Righi et al. (2017b) we specifically derived the expected significance of a possible detection by IceCube of Mkn 421, obtaining a significance of 3$\sigma$ after 8 years (although these estimates are base on the somewhat extreme assumption that BL Lacs account for the entire neutrino diffuse flux). 
The lack of any excess around the position of these two sources, together with the possible observation of a neutrino emission by TXS 0506+056 source (not a HSP), bring us to ponder about the photon component involved on the photo-meson reactions. In Tavecchio et al. (2014) and Tavecchio et al. (2015), the photons produced in the external and slow sheath of the jets is thought to play a role on the neutrino productions. This scenario is applied to the high-energy emitting BL Lacs, those sources in which there are the strongest indications supporting the presence of the spine-layer structure. The problems with Mkn 421 and Mkn 501 lead us to propose that the radiatively inefficient radiation flow can provide a radiation field that would favour LSP sources as neutrino emitters and would disfavour ISP and HSP objects (Righi et al. 2018).

A possible continuation of the study described in this paper could be the extension to the BL Lac objects of the Third Catalog of Hard Fermi-LAT Sources (3FHL; Ajello 2017), which contains the sources detected in 7 years above $10$ GeV by Fermi. This catalogue is composed of $\sim 50\%$ HSP and $\sim 50\%$ ISP+LSP. Table 7 shows the spatial correlation with the same sample of neutrino events and the BL Lacs of the 3FHL catalogue. A in-depth study of the SED and the light curve of these sources will be pursued.
 

\begin{table}
\centering
\begin{tabular}{l|cc|cc|c}
\hline
Source name & $\alpha$ & $\delta$  & $z$ & Class & $\nu$\\
& (J2000) & (J2000) & & & ID\\
\hline
\multicolumn{6}{c}{Single 3FHL BL Lac inside the angular uncertainty of the HESE events} \\
\hline
PKS1101-536                             & 165.967 & -53.950  & -      &LSP   &  4$^a$\\
1RXSJ094709.2-254056            & 146.789 & -25.683 & -      & -        &  46$^b$\\
NVSSJ173146-300309              & 262.945 & -30.052  & -      & -        &  14$^a$\\
\hline
\multicolumn{6}{c}{3FHL BL Lac with a distance max of 2.5$^\circ$ from a $\nu_\mu$} \\
\hline
\textbf{NVSSJ140450+655428} & 211.206 & 65.908 & 0.363 &  HSP  & 47$^a$\\
\textbf{4C+41.11}                       & 65.983   & 41.834 & -         &  LSP  & 13$^a$\\
\textbf{MG1J021114+1051}       & 32.804   & 10.859 & 0.200 &  ISP & 23$^b$\\
\textbf{TXS 0506+056}              & 77.358   &  5.693  & 0.336 &  ISP & *$^c$\\
PMNJ2227+0037                      & 336.992 &  0.618   & 2.145 & ISP  & 44$^b$\\
PMNJ0152+0146                      & 28.165   &  1.788   & 0.080 & HSP& 1$^d$\\
MG3J225517+2409                  & 343.779 &  24.187 & -         & LSP &  3$^d$\\
RXJ1533.1+1854                      & 233.296 & 18.908 & 0.307 & HSP & 12$^d$\\
RXJ2030.8+1935                     & 307.738 & 19.603  & -         &  -      &  5$^d$\\
1ES0229+200                          & 38.202 & 20.288    & 0.140 &  HSP & 16$^d$ \\

\hline
\end{tabular}
\label{Table3fhl}
\caption{List of candidates neutrino sources of 3FHL. Bold face characters identify the sources of 3FHL studied in this paper. The neutrino ID is taken from: $a$: Aartsen et al. 2014, $b$: Aartsen et al. 2015, $c$: Aartsen et al. 2018, $d$: Aartsen et al. 2016. We show the redshift reported in NED.}
\end{table}

\section*{Acknowledgments}
The authors acknowledge contribution from the grant INAF CTA--SKA ``Probing particle acceleration and $\gamma$-ray propagation with CTA and its precursors''. Part of this work is based on archival data and  on--line services provided by the ASI SDC. We acknowledge financial contribution from the agreement ASI-INAF I/037/12/0.
This work made use of data supplied by the UK Swift Science Data Centre at the University of Leicester. We thank the referee for his/her suggestions that helped us to improve the paper.

\end{document}